\documentstyle[prl,aps,epsf]{revtex}
\begin{document}
\draft
\wideabs{
\title{Vacuum--Stimulated Raman Scattering based on\\ Adiabatic 
Passage in a High--Finesse Optical Cavity}
\author{M. Hennrich, T. Legero, A. Kuhn, and G. Rempe}
\address{Max-Planck-Institut f\"ur Quantenoptik, 
Hans-Kopfermann-Str.\,1, D--85748 Garching, Germany}
\date{Phys. Rev. Lett. 85, 4872-4875 (2000); Received 21 August 2000}
\maketitle
\begin{abstract}
We report on the first observation of stimulated Raman scattering 
from a $\Lambda$--type three-level atom, where the stimulation is realized 
by the vacuum field of a high-finesse optical cavity. 
The scheme produces one intracavity photon by means of an adiabatic 
passage technique based on
a counter-intuitive interaction 
sequence between pump laser and cavity field.
This photon leaves the cavity through the less-reflecting mirror.
The emission rate shows a characteristic 
dependence on the cavity and pump detuning, and the observed spectra 
have a sub-natural linewidth.
The results are in excellent agreement with 
numerical simulations.
\end{abstract}
}

\narrowtext

In the last few years,  interesting proposals on the generation of non-classical 
states of light in optical cavities \cite{Parkins,Lange00} and on 
the controlled generation of single photons from such cavities 
\cite{Law,Kuhn99} were made. All these schemes are based on 
a technique known as {\sc Stirap} (stimulated Raman scattering involving adiabatic passage)
\cite{Stirap,Kuhn98} or a variant thereof, and incorporate the time 
dependent
interaction of an atom with the field mode of an optical cavity. 
The operation principle is related to that of a Raman laser 
\cite{Becker87},
with the difference that now a single atom interacts with an empty 
cavity mode.
Other schemes for the preparation of Fock states are based on
vacuum Rabi oscillations or, more generally, $\pi$-pulses in a 
two-level atom. In these cases, the need of a long-lived excited atomic state 
restricts experiments to the microwave regime \cite{Haroche,Walther}, 
where the photon remains stored in a high-Q cavity.

Here, we report on the experimental realization of 
an excitation scheme that allows one to emit a visible
photon into a well defined mode
of  an empty cavity. This photon then leaves the cavity in a known 
direction.
Our method is  based on the single-photon 
generation scheme discussed in  \cite{Kuhn99}. It relies on  {\sc Stirap} 
\cite{Stirap,Kuhn98}, but instead of using two delayed laser 
pulses, we have only one exciting pump laser, combined with a strong 
coupling of a single atom to a single cavity mode 
\cite{Mabuchi96,Munstermann99}. This strong coupling 
 induces the anti-Stokes transition of the Raman 
process. 

\begin{figure}
\centerline{\epsfxsize=5.5cm\epsfbox{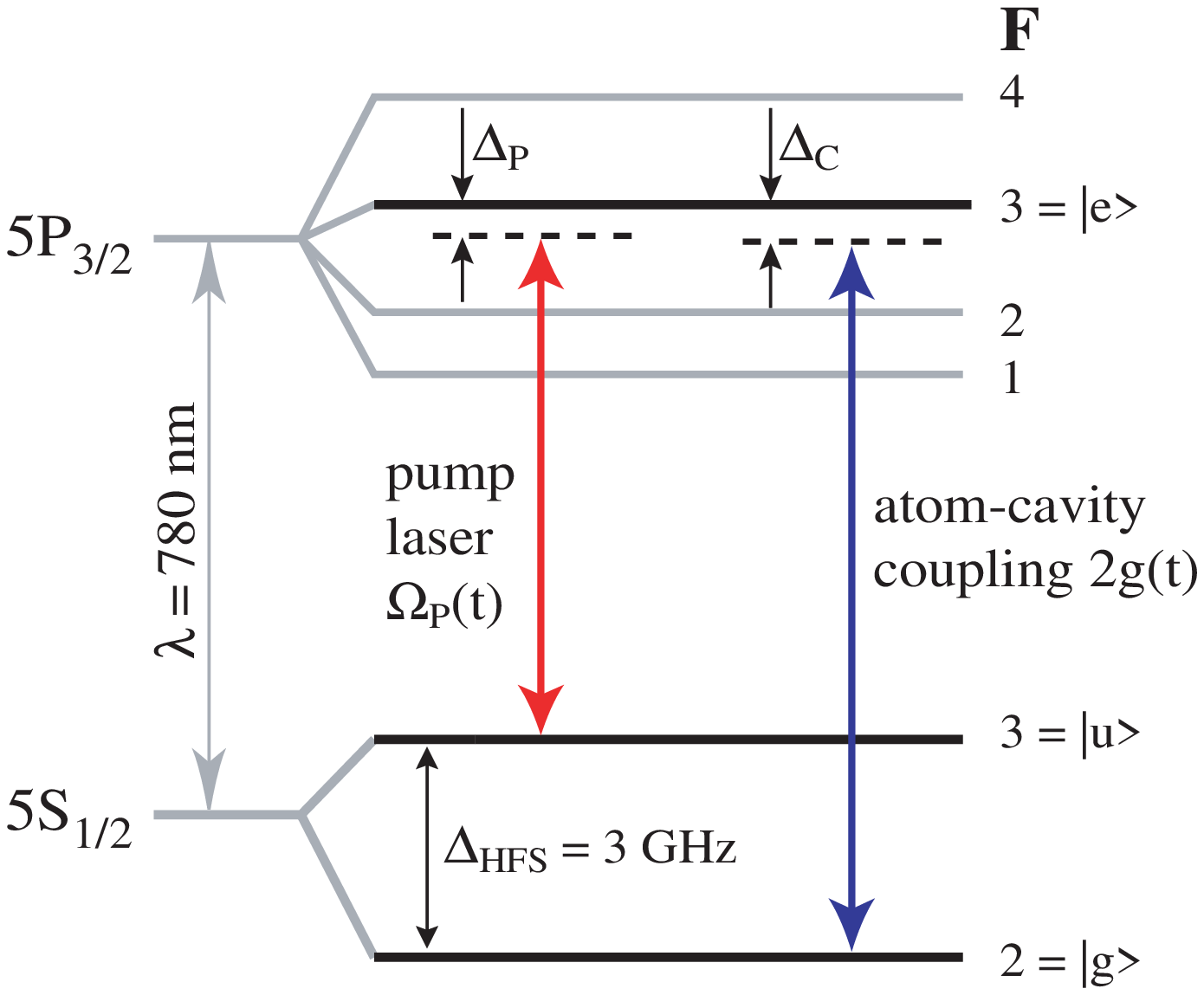}} \vspace{3mm} 
\caption{Scheme of the relevant levels, transitions, and detunings
of the $^{85}$Rb--atom coupled to the pump laser and the cavity.\label{level31}}
\end{figure}

Figure \ref{level31} depicts the excitation scheme for the 
$^{85}$Rb--atoms  used in our experiment. A $\Lambda$-type 
three-level scheme is realized by the two $5S_{1/2}$ hyperfine 
ground states $F=3$ and $F=2$, which we
label  $|u\rangle$ and $|g\rangle$, respectively. The $F=3$ 
hyperfine level of the electronically excited state $5P_{3/2}$ forms
the intermediate state, $|e\rangle$.
The atom interacts with a single-mode of an optical cavity, with states 
$|0\rangle$ and $|1\rangle$ denoting zero and one photon in the mode,
respectively. The cavity resonance frequency, $\omega_{C}$, is close to the 
atomic transition frequency between states 
$|e\rangle$ and $|g\rangle$, but far-off 
resonance from the $|e\rangle$ to $|u\rangle$ transition. Hence, 
only the product states $|e,0\rangle$ 
and $|g,1\rangle$ are coupled by the cavity. For this transition, 
the  vacuum Rabi frequency,
\begin{equation}
2g(t)=2g_{0}\exp\left(-\left(\frac{t\, v}{w_{C}}\right)^2\right),
\end{equation}
is time dependent since the atom moves with velocity $v$ across the waist 
$w_{C}$ of the Gaussian cavity mode. Its peak amplitude is given by the atom-cavity 
coupling coefficient at an antinode, $g_{0}$.

%%%%%%%%%%% figure 1 %%%%%%%%%%%%%%%%%%%%%%%%%%%%%%%%%%%
%%%%%%%%%%% figure 1 %%%%%%%%%%%%%%%%%%%%%%%%%%%%%%%%%%%

In addition to the interaction with the cavity mode, the atom is 
exposed to a pump laser beam which crosses the cavity 
 axis at right angle. This beam is placed slightly downstream in the path of the 
 atoms (by $\delta_{x}$ with respect to the cavity axis) and has a 
waist $w_{P}$, therefore causing a time dependent Rabi frequency 
\begin{equation}
\Omega_{P}(t)=\Omega_{0}\exp\left(-\left(\frac{t\, v - 
\delta_{x}}{w_{P}}\right)^2\right).
\end{equation}
The pump frequency is  near
resonant with the transition between $|u,0\rangle$ and $|e,0\rangle$, 
thereby coupling these states. 

In a  frame rotating with the cavity frequency and the pump laser 
frequency, the Hamiltonian is given by
\begin{eqnarray}
H(t)&&= \hbar [ \Delta_{P}|u\rangle\langle u| + \Delta_{C}|g\rangle\langle  g|\\
&& + g(t) (|e\rangle\langle g| a + a^\dag |g\rangle\langle e|) + \frac{1}{2}\Omega_{P}(t) 
(|e\rangle\langle u| + |u\rangle\langle e| )].\nonumber
\end{eqnarray}
Here,  $\Delta_{C}$ and $\Delta_{P}$ denote the detunings 
of the cavity and the pump beam from their respective atomic 
resonances, and $a$ and $a^\dag$ are the  annihilation and creation operators 
of 
the cavity field. The pump beam is treated semiclassically.
On Raman resonance, i.e. for $\Delta_{C}=\Delta_{P}$, one of the eigenstates of this interaction Hamiltonian 
reads
\begin{equation}
|a^0(t)\rangle = \frac{2 g(t) |u,0\rangle - \Omega_{P}(t) |g,1\rangle}{\sqrt{4 
g^2(t) + \Omega_{P}^2(t)}}.\label{anull}
\end{equation}
This is a dark state without any contribution from the electronically excited 
level $|e,0\rangle$. Therefore losses due to spontaneous emission cannot occur, 
provided the state vector of the system,  
$|\Psi\rangle$, follows $|a^0\rangle$  throughout the 
Raman excitation.

%%%%%%%%%%% figure 2 %%%%%%%%%%%%%%%%%%%%%%%%%%%%%%%%%%%
\begin{figure}
\centerline{\epsfxsize=6.5cm\epsfbox{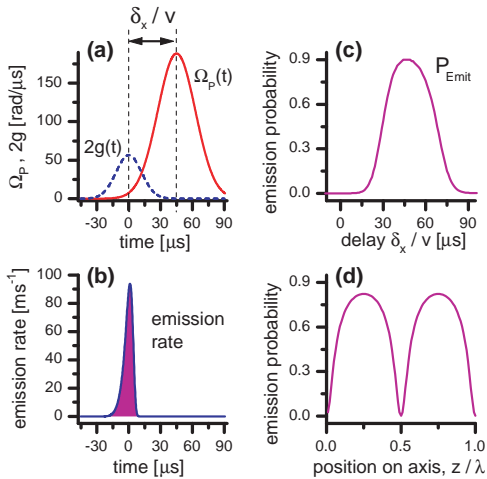}} \vspace{3mm} 
\caption{Simulation of a resonant atom-cavity interaction sequence for a cavity decay 
constant,  $2\kappa = 2\pi\times 2.5\,$MHz,  an atomic decay constant of $\Gamma=2\pi\times 
6\,$MHz, and atoms travelling at  $v=2\,$m/s.
{\bf(a)} $\Omega_{P}(t)$ and  $2 g(t)$  shown for experimental 
amplitudes and waists,  $w_{C}=35\,\mu$m and $w_{P}=50\,\mu$m.
{\bf(b)} photon emission rate for a delay of  
$\delta_{x}/v=45\,\mu$s. The integral of the rate yields a total photon 
emission probability, $P_{Emit}$, of $90\%$.
{\bf(c)}  $P_{Emit}$  as a function 
of the delay, $\delta_{x}/v$,  between cavity and pump interaction.
{\bf(d)}  $P_{Emit}$  as a function of the atomic position on the 
cavity axis for a delay of $\delta_{x}/v=35\mu$s.
\label{oneshot}}
\end{figure}
%%%%%%%%%%% figure 2 %%%%%%%%%%%%%%%%%%%%%%%%%%%%%%%%%%%

The atom is prepared in state $|u\rangle$ 
before it enters the empty cavity, i.e. atom and field start in state $|u,0\rangle$. 
Since the pump beam is displaced by $\delta_x$ with respect to the cavity 
axis, the atom is subject to a counter-intuitive delayed pulse 
sequence, i.e. due to the initial condition $2g \gg 
\Omega_{P}$, the evolution starts with $\langle\Psi|a^0\rangle=1$. The subsequent interaction
with the pump beam leads to $\Omega_{P}\gg 2g$, which implies the  evolution of
$|a^0\rangle$ into state $|g,1\rangle$. Provided the state vector 
$|\Psi\rangle$ is able to follow,  the system is
transferred to  $|g,1\rangle$, and a photon is placed in the cavity mode. 
Since this photon is emitted with 
the cavity energy decay rate, $2\kappa$, the empty cavity state,
$|g,0\rangle$, is 
finally reached and the  atom-cavity system decouples from any further interaction.

This simple excitation scheme relies on three conditions. First, the 
detunings of the cavity, $\Delta_{C}$, and of the pump pulse, 
$\Delta_{P}$, must allow a Raman transition, i.e.
\begin{equation}|\Delta_{C}-\Delta_{P}| < 2 \kappa.\end{equation}
Second, the condition for $|\Psi\rangle$ adiabatically following 
$|a^{0}\rangle$ must be met \cite{Kuhn99,Stirap},
\begin{equation}
(2g_{0} w_{C} / v \, ,\, \Omega_{0} w_{P} / v )\gg 1.
\end{equation}
Third, either the interaction time must be significantly longer than $(2\kappa)^{-1}$  
to allow the emission of the photon before it is  
reabsorbed by the atom
due to coherent population return, \cite{Kuhn99,Kuhn98},
or alternatively, the interaction with the pump beam must 
be strong when the atom leaves the cavity to avoid this reverse 
process.

%%%%%%%%%%% figure 3 %%%%%%%%%%%%%%%%%%%%%%%%%%%%%%%%%%%
\begin{figure}
\centerline{\epsfxsize=6cm\epsfbox{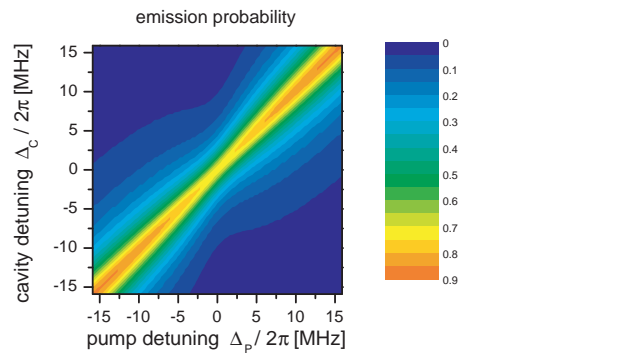}} \vspace{3mm} 
\caption{Photon emission probability as a function of cavity and pump 
detuning, calculated for a pulse delay of $\delta_x/v=35\,\mu$s and 
the parameters of Fig.\,\ref{oneshot}. The chosen delay gives the best 
fit with the experimental data shown in Fig.\,\ref{results}. \label{graymap}}
\end{figure}
%%%%%%%%%%% figure 3 %%%%%%%%%%%%%%%%%%%%%%%%%%%%%%%%%%%

A numerical simulation  for a 
single atom crossing the cavity is shown in Fig.\,\ref{oneshot}. To 
include the cavity-field decay rate, $\kappa$, and the spontaneous 
emission rate of the atom, $\Gamma$, we have 
employed the density-matrix formalism described in \cite{Kuhn99}. For 
the resonant situation, $\Delta_{P}=\Delta_{C}=0$ shown here, the 
total emission probability, $P_{\mbox{\tiny Emit}}$,  is expected to reach 90\%. 
For the considered waists and amplitudes, Fig.\,\ref{oneshot}(c) shows 
that $P_{\mbox{\tiny Emit}}$ reaches its maximum for $\delta_x/v=45\,\mu$s. Note also that
$P_{\mbox{\tiny Emit}}$ is vanishingly small if the interaction with the pump beam 
coincides or precedes the interaction with the cavity mode. 
Fig.\,\ref{oneshot}(d) shows $P_{\mbox{\tiny Emit}}$ as a function of 
the atom's position on the cavity axis for the delay realized in the 
experiment. Due to the standing wave mode 
structure, the emission probability is zero at the nodes, and shows 
maxima at the antinodes. Since the dependence of $P_{\mbox{\tiny 
Emit}}$ on the position-dependent coupling constant, $g$, is highly nonlinear and 
saturates for large $g$, the gaps around the nodes are much narrower 
than the plateaus surrounding the antinodes. 

%%%%%%%%%%% figure 4 %%%%%%%%%%%%%%%%%%%%%%%%%%%%%%%%%%%
\begin{figure}
\centerline{\epsfxsize=5.5cm\epsfbox{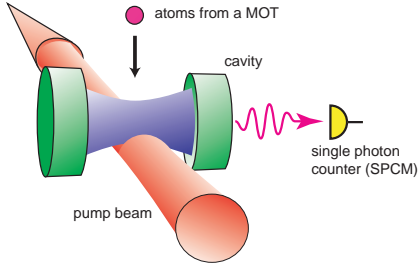}} \vspace{3mm} 
\caption{Sketch of the experimental setup. 
The pump beam is displaced with respect to the cavity mode. \label{setup}}
\end{figure}
%%%%%%%%%%% figure 4 %%%%%%%%%%%%%%%%%%%%%%%%%%%%%%%%%%%

Fig.\,\ref{graymap} depicts  the case where 
$\Delta_{P}\neq\Delta_C$. It is obvious that $P_{\mbox{\tiny Emit}}$ 
is close to unity if the excitation is Raman resonant $(\Delta_{P}=\Delta_C)$. 
However, for the delay $\delta_x/v=35\,\mu$s chosen here, a smaller signal 
is expected for $\Delta_{P}=\Delta_{C}=0$, since the waist of the 
pump, $w_{P}$, is larger than $w_{C}$, and resonant excitation of the 
atom prior to the interaction with the cavity mode cannot be neglected.

To realize the proposed scheme, we have chosen the setup  
sketched in Fig.\,\ref{setup}. A  
cloud of $^{85}$Rb atoms is prepared in the $5S_{1/2},\,F=3$ state and 
released from a magneto-optical 
trap (MOT) at a temperature of  $\approx 10\,\mu$K. A small fraction 
(up to 100 atoms) falls through a stack of apertures  
and enters the mode volume of an optical cavity at a speed of 2\,m/s. 
The cavity is composed of two mirrors with a curvature of 50\,mm and a 
distance of 1\,mm. The waist of the TEM$_{00}$ mode is 
$w_{C}=35\,\mu$m, and in the antinodes the coupling 
coefficient is $g_{0}=2\pi\times 4.5$\,MHz. The finesse of 61\,000 
corresponds to a 
linewidth $2 \kappa = 2\pi\times 2.5\,$MHz (FWHM), which is significantly smaller 
than the natural linewidth of the $^{85}$Rb atoms.
While one cavity mirror is highly reflective 
($1-R=4\times10^{-6}$),  the transmission of the other  is 
$25\times$ higher to emit the photons in one direction only. 
A single-photon counting module (SPCM) with a quantum 
efficiency of 50\% is used to detect them.

A reference laser is used to stabilize the cavity close to resonance with the $5S_{1/2}, F=2 
\longleftrightarrow 5P_{3/2}, F=3$ transition with a lock-in technique.
However, since an empty cavity is needed for the experiment, this laser is 
blocked  3.7\,ms before the atoms enter the cavity.

The pump beam is close to resonance with the $5S_{1/2}, F=3 
\longleftrightarrow 5P_{3/2}, F=3$ transition and crosses the cavity 
transverse to its axis. This laser is focussed to a waist of 
$50\,\mu$m and has a power of $5.5\,\mu$W, which corresponds to a 
peak Rabi frequency $\Omega_{0} = 2\pi\times 30\,$MHz.

The desired counter-intuitive pulse sequence for {\sc Stirap}
is realized by time of flight. The atoms first enter the cavity mode  and therefore 
experience a strong coupling on the anti-Stokes transition, whereas the 
interaction with the pump beam is delayed, since it crosses the 
cavity mode slightly downstream. This delay has 
been optimized to achieve a high flux of photons leaving the cavity. 

%%%%%%%%%%% figure 5 %%%%%%%%%%%%%%%%%%%%%%%%%%%%%%%%%%%
\begin{figure}
\centerline{\epsfxsize=6.4cm\epsfbox{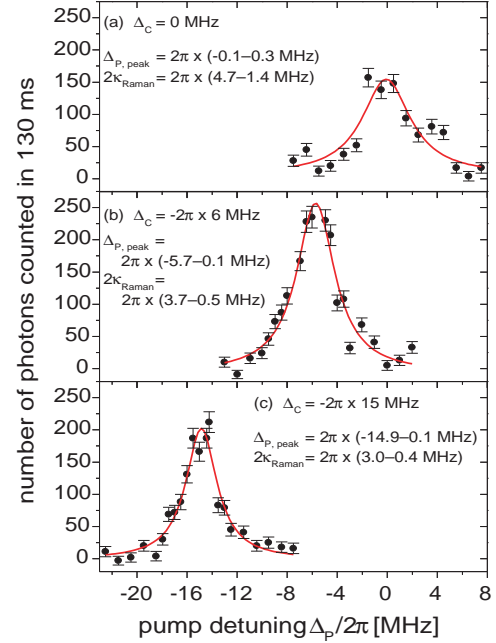}} \vspace{3mm} 
\caption{Number of photons from the cavity as a
function of the pump laser detuning, $\Delta_{P}$, for three different 
cavity detunings. The solid lines are Lorentzian fits to 
the data. \label{results}}
\end{figure}
%%%%%%%%%%% figure 5 %%%%%%%%%%%%%%%%%%%%%%%%%%%%%%%%%%%

Figure \ref{results}(a) shows the number of counted photons emerging 
from the cavity as a function 
of the pump pulse detuning, $\Delta_{P}$, in case of a resonant 
cavity, $\Delta_{C}=0$. The detunings of the cavity and the pump laser 
are both adjusted  by means of acousto-optic modulators. 
To register the data, the MOT has been 
loaded and dropped across the cavity 50 times. The atom cloud needs 
6.5\,ms (FWHM) to cross the cavity mode, and within this interval, the photons 
emerging from the cavity are measured by the SPCM 
and recorded by a transient digitizer during 2.6\,ms with a time 
resolution of 25\,MHz. Therefore, the signal is observed for a total 
time of 130\,ms. Due to the dark count rate of 390\,Hz of the SPCM, 
the total number of dark counts in the interval is limited to $51\pm 7$.  

In the resonant case, one expects a small probability for atomic 
excitation. This could lead to a small but cavity enhanced spontaneous 
emission into the cavity mode, as has been shown previously 
\cite{spontan}. Our numerical simulation shows that an excited atom at 
the antinode
emits into the resonant cavity mode with a probability that can be as high 
as  26\%, indicating that even in this case most of the
 spontaneously emitted photons are lost 
in a random direction. This loss explains the smaller 
peak emission rate with respect to the off-resonant cases discussed 
below.
Note that the cavity mode covers only a small solid angle of $\approx 4\pi\times 
2.6\times 10^{-5}$ srad, therefore the calculated spontaneous emission 
rate into the cavity is enhanced by a factor $10^4$. 
However, the linewidth is 
sub-natural, and therefore the observed signal can not be attributed 
to an excitation by the pump beam followed by enhanced spontaneous 
emission.

This is even more evident if the cavity is detuned   (Fig.\,\ref{results}(b,c)).
The emission peak is pulled away from the atomic 
resonance, following the Raman resonance 
condition, $\Delta_{P}=\Delta_{C}$. Such a displacement proofs that
 the  light emission is not the result of a pump 
transition followed by enhanced spontaneous emission into the cavity.
Moreover, $\Delta_{P}$ is too high for an electronic excitation of the 
atoms. Therefore, the far-out reaching wings of the
pump beam do not excite the atoms prior to their interaction with the cavity 
mode any more. The losses vanish, and the peak photon emission probability is 
higher than for the resonant case. Note also that
the observed linewidth is much smaller than the natural linewidth, 
$\Gamma = 2\pi\times 6\,$MHz, of the atom. For $\Delta_{C}=-2\pi\times 15\,$MHz, 
the line is only 3\,MHz wide and  approaches the 
linewidth $2\kappa=2\pi\times 2.5\,$MHz of the
cavity, which also limits the width of the Raman transition, since 
$2\kappa$ is the decay rate of the final state, $|g,1\rangle$. 

In our discussion, we have assumed that the atoms interact
with the cavity one-by-one. This is 
justified according to the following estimation: A  mechanical slit restricts the
atom's maximum distance from the 
cavity axis to $\pm 50\,\mu$m. The spatial variation of $g$ along  
(Fig.\,\ref{oneshot}(d)) and perpendicular to the cavity axis reduces the average 
emission probability to 37\% per atom crossing the slit and the pump beam. 
Due to the low quantum efficiency of the SPCM and unavoidable cavity 
losses, only about 40\% of the generated photons are detected. Therefore 
the maximum measured rate of 230\,events/130\,ms corresponds to a 
generation rate of 4.4  photons/ms, and at least 12 atoms/ms are needed to explain this signal.
Since the photon generation takes $12\,\mu$s 
(FWHM, Fig.\,\ref{oneshot}(b)), the probability that a second
 atom interacts with the cavity simultaneously is 14\%. This is 
small and, hence, negligible.

All observed features are in excellent agreement with our simulation, and
we therefore conclude that the photon emission is caused by a 
vacuum-stimulated Raman transition, i.e. the coupling to the cavity, $g(t)$, and 
the Rabi frequency of the pump laser, $\Omega_{P}(t)$, are both high 
enough to assure an adiabatic evolution of the system, thus forcing the state
vector $|\Psi\rangle$ to follow the dark state $|a^0\rangle$ throughout the interaction.
Loss due to spontaneous emission is suppressed, and the 
 photons are emitted into a single 
mode of the radiation field with well determined frequency
and  direction.

The scheme can be used
to generate single, well characterized photons on demand, provided 
the Raman excitation is performed in a controlled, triggered way. 
In contrast to 
other single-photon sources \cite{singlephoton}, these photons will 
have a narrow bandwidth and a directed emission. 
Finally, we state that the photon generation process depends on the 
initial state of the atom interacting with the cavity. If the atom is 
prepared in a 
superposition of states $|g,0\rangle$ and $|u,0\rangle$ prior 
to the interaction, this state will be mapped  onto the emitted 
photon. 
A second atom placed in another cavity  could 
 act as a receiver, and with the suitable pump pulse sequence applied  to 
 the emitting and the receiving atom, a quantum teleportation of 
the atom's internal state could be realized \cite{Cirac97}.

This work was partially supported by the  focused research program ``Quantum 
Information Processing'' of the Deutsche Forschungsgemeinschaft
and  the QUBITS project of the IST program of the European Union.

%%%%%%%%%%%%%%%%%%%%%%%%%% FIGURES %%%%%%%%%%%%%%%%%%%%

\end{document}